# Detecting the Glint of Starlight on the Oceans of Distant Planets

Darren M. Williams

School of Science; Penn State Erie, The Behrend College

4205 College Drive

Erie, Pennsylvania 16563-0203

Phone: 814-898-6008

Fax: 814-898-6213

dmw145@psu.edu

Eric Gaidos

Dept. of Geology and Geophysics

University of Hawaii

Honolulu, Hawaii 96822

gaidos@hawaii.edu

41 pages

7 Figures

**Running Head**

Glint of Starlight on the Oceans of Distant Planets

**Principal Correspondence**

Darren M. Williams

School of Science; Penn State Erie, The Behrend College

5091 Station Road

Erie, Pennsylvania 16563-0203

Phone: 814-898-6008

Fax: 814-898-6213

dmw145@psu.edu

# Abstract

We propose that astronomers will be eventually be able to discriminate between extrasolar Earth-like planets with surface oceans and those without using the shape of phase light curves in the visible and near-IR spectrum. We model the visible light curves of planets having Earth-like surfaces, seasons, and optically-thin atmospheres with idealized diffuse-scattering clouds. We show that planets partially covered by water will appear measurably brighter near crescent phase (relative to Lambertian planets) because of the efficient specular reflection ("glint") of starlight incident on their surfaces at a highly oblique angle. Planets on orbits within 30º of edge-on orientation (50% of all planets) will show pronounced glint over a sizeable range of orbital longitudes, from quadrature to crescent, all outside the glare of their parent stars. Also, water-covered planets will appear darker than a Lambertian disk near full illumination. Finally, we show that planets with a mixed land/water surface will polarize the reflected signal by as much as 30-70%. These results suggest several new ways of directly identifying water on distant planets.

# Keywords

# Introduction

Astronomers are discovering extrasolar planets at a remarkable pace. Nearly all of the more than 200 or so planets found in the last decade (Butler *et al*. 2006) are comparable in mass to Jupiter and have been identified from the reflex motion of stars accelerated by their planet's gravity. The planets themselves are too faint ($10^{-9}$– $10^{-8}$ parent-star luminosity) to be directly imaged at present, but proposed space telescopes such as TPF-C (Traub *et al.* 2006), TPF-I (Des Marais *et al*. 2002; Lawson *et al*. 2006), and Darwin (Kaltenegger and Fridlund 2005) should be able to resolve star-planet pairs and null the light of the parent star to a degree that will enable astronomers to directly observe planets as small as Earth, locate the planets relative to the habitable zone (Kasting et al. 1993)**,** estimate their surface temperatures, and look for water vapor and molecular biomarkers in their atmospheres (Des Marais *et al*. 2002; Kaltenegger and Fridlund 2005). Also, visible and infrared light curves over an entire orbit can be assembled from observations of an extrasolar planet at multiple epochs.

There are many factors that can affect the apparent brightness of an Earth-like planet in reflected light ($\lambda < 1$ μm) as it orbits its star. A planet may change color and vary in brightness by a few hundred percent as it rotates in hours to days (Ford *et al*. 2001). Desert-covered land masses and bright clouds account for the bulk of the diurnal variation on Earth and can be seen in earthshine reflected by the Moon (Goode *et al*. 2001; Qiu *et al*. 2003). Cloud-generating weather patterns on an extrasolar planet could be evidence for an active hydrologic cycle and the presence of surface water. However, such diurnal variations will be difficult to detect on planets with rotational periods under a few days because a TPF-class instrument may require 5-15 days to effectively observe them (Brown 2005; Traub *et al.* 2006; Also see Fig. 1).

Changes in planet brightness on an orbital or seasonal timescale would be apparent in a photometric light curve assembled from 10-15 observations, each integrating over many rotations of the planet. Seasonal changes in reflectivity would result from a non-zero obliquity or an eccentric orbit, and the associated freezing and thawing of snow and ice. Warmer, ice-free planets might also exhibit measurable seasonality from changes in cloud cover, hemispheric blooms of land plants (Montañéz-Rodríguez *et al*. 2006) or oceanic algae (Knacke 2003), or the lofting and deposition of particles by winds, as occurs on Mars (Christensen 1988). Of all these possibilities, seasonal whitening of a planet by clouds or snow (albedo $A = 0.75 - 0.95$) will yield the greatest variability since the albedo of most unfrozen surfaces, including water under average illumination, is less than 25%.

Seasonal albedo fluctuations will vary from planet to planet because they depend on a planet's obliquity and orbital inclination with respect to the observer, as well as the many other surface details affecting climate (e.g., size and location of continents, and the thermal time-constant and heat-transport efficiency of the atmosphere-ocean system). In addition to the albedo cycle, all planets with orbits inclined relative to the plane of the sky will exhibit a repeating cycle of phases that depends in a known way (and a *measurable* way through astrometry) on orbital inclination $i$ and circumstellar orbital longitude $\theta$ through the relation

$$f_A = (1 - \cos\theta \, \sin i)/2. \qquad (1)$$

Here, $f_A$ is the fraction of the projected disk that is illuminated by the parent star. Planets with small inclinations (approximately face-on orbits) will demonstrate flat phase light-curves, whereas planets on steeply inclined orbits will exhibit a sinusoidal phase variation as indicated in Fig. 2. A planet is in gibbous phase near the center of the figure and in crescent phase near the left and right edges.

Not all phases in the cycle will be observable because of limits to angular resolution and contrast ratio. Orbital inclination and star-planet separation $a$ determine the edges of the viewing window when the star-planet pair can be resolved. For nearly edge-on orbits ($i > 75º$), a planet 10 parsecs away and 1.0 AU from its parent star will be observable ~67% of the time (i.e., between $\theta \sim 30\text{-}150º$ and $\theta \sim 210\text{-}330º$) when it is outside the inner working radius $4\lambda/D \sim 0.057``$ for an 8-meter telescope observing at $\lambda = 550$ nm. The phase range for planet detection is somewhat wider for less-inclined orbits, but narrower for planets closer to their stars. According to Fig. 2, it will be impossible to detect thin-crescent planets ($f_A < 0.25$) at any inclination if they are within 0.66 AU of the parent star. Thus, this approach is applicable only to planets in the habitable zones around G and F stars where they are most likely to be resolved (cf. Fig. 2)

A reflected light curve also contains information about the scattering properties of the surface, independent of any seasonal changes. Planets with water will reflect light toward the observer more efficiently in crescent phase than in gibbous phase because of the higher reflectance at low incident angles. This glint from water will make a planet appear anomalously bright in crescent phase compared to diffuse-scattering surfaces observed in the same geometry. Light reflected from water will also impart some polarization to the disk-averaged signal, which might be measurable under idealized (i.e., optically-thin, cloud-free) atmospheric conditions. Here we investigate whether specular reflection from water can be identified by either photometric or polarimetric measurements made with a TPF-like telescope.

## Model Description

Most planetary surfaces are diffuse scatters and their scattering properties can be approximated by an isotropic (Lambertian) law; the notable exceptions being smooth clean ice, optically thin clouds, and liquids such as water, or ethane as on Titan (Campbell *et al.* 2003; Lorenz 2003; Mitri *et al.* 2007). A liquid ocean on a terrestrial planet in the habitable zone of a star is most likely to be made of water based on its thermal properties, as well as the abundance of water in the Solar System and in the protoplanetary disks of young stars (Eisner 2007). Here we consider a planet of radius $R_{\mathrm{p}}$ at a distance $a$ from a star of luminosity $L_*$ that reflects power per solid angle $F_{\mathrm{p}}$ in the direction of Earth equal to

$$F_{\mathrm{p}} = \frac{L_*}{4\pi a^2} \frac{f_A \pi R_p^2}{2\pi} \left[ f_{diff} A_{diff} + \frac{f_{spec} A_{spec}}{f_\Omega} \right] \text{ (W sr}^{-1}\text{)},$$

$$= F_0 f_A \left[ f_{diff} A_{diff} + \frac{f_{spec} A_{spec}}{f_\Omega} \right] \text{ (W sr}^{-1}\text{)}, \qquad (2)$$

where all of the star and planet parameters have been swept into the parameter $F_0$, and $f_A$ is given by Eqn. 1. The terms in brackets are total diffuse light and specular-reflected light, with $f$ denoting area fraction and $A$ representing surface albedo. Specularly-reflected light from water is scattered into a solid angle $f_\Omega 2\pi$ ($f_\Omega = 1$ for Lambertian surface; $f_\Omega < 1$ otherwise) equal to the area angle of the star on the planet, $2\pi$ (1 −

cos($2R_*/a$)) $\approx 2\pi \times 10^{-5}$ sr for the Sun on the Earth. Glint from a wavy ocean surface arrives from an area considerably wider than this (some 30° wide covering 0.214 sr in because the light is reflected from an aggregate of wave surfaces each having a slope and orientation such that the geometry for specular reflection is satisfied. Light from the center of the glint spot reflects from waves of lesser slope than near the spot edges, where the reflection angle relative to normal incidence equals the maximum wave slope expected for a given surface wind.

Fig. 3a; comparable to what is seen in satellite images of the Earth – see Fig. 4a)

The surfaces of our model planets are described by a $180 \times 90$ Mercator grid with each

2°x 2° pixel assigned one of the three surface types: unfrozen land, snow and ice, and

water. Continental geography is set to that of the present Earth (except for simulations

with the entire planet covered by water), and the area covered by snow and ice is

updated monthly using surface albedo maps generated from three-dimensional climate

simulations of the Earth (Williams and Pollard 2003). The probability distribution of

wave surfaces is approximated by a gaussian (Cox and Munk 1954; Vokroughlický and

Farinella 1995) of the form

$$\ln(p_{wav}) \propto -\tan^2\alpha/2\sigma^2, \qquad (3)$$

where $\alpha$ is the wave tilt and

$$\sigma^2 = 0.003 + 0.00512 \ (v/1 \ \text{m sec}^{-1}), \qquad (4)$$

with v equal to the mean surface wind speed (Haltrin 2002). We assume a maximum wave tilt $\alpha = 25°$ (Vokroughlický and Farinella 1995) and set v = 10 m sec$^{-1}$, which sets the surface fraction covered by diffusely-scattering sea foam

$$f_{foam} = 0.0000125 \ (v/1 \ m/sec)^{3.3} = 2.5\% \qquad (5)$$

(Haltrin 2002). Every 2 x 2° pixel is assigned an identical cloud fraction between 0 - 50%, the range expected for Earth-like planets with clouds formed through convection. Optically-thick water clouds are assumed to scatter light isotropically with an albedo of 60%. A more realistic cloud parameterization would include anisotropic scattering described by a Henyey-Greenstein (or similar) phase function with forward and back scattering lobes, as has been used to successfully model light reflected from the cloud decks of Venus (Mallama 2006), and Jupiter and Saturn (Dyundina 2005). Here we ignore the effects of anisotropic cloud scattering because our primary focus is to show how water contributes to the signal reflected by the surface. However, we acknowledge that forward scattering by clouds (as well as molecular scattering by the atmosphere) may obfuscate the interpretation of light received from planets in the crescent phase of their orbits. We will address this point further below.

Rewriting Eqn. 2 for planets covered by a mix of diffuse-reflecting surfaces plus water yields:

$$F_{\mathrm{p}} = F_0 f_A \left[ f_{cld} A_{cld} + f_{\ln d} A_{\ln d} + f_{ice} A_{ice} + \frac{f_{wtr} \, p_{wav} A_{wtr}}{f_\Omega} \right] \ (\mathrm{W \ sr\text{-}1}). \qquad (6)$$

The diffuse signal is assumed to come from the combined contribution of water clouds and sea foam (both assumed here to have the same albedo: $A_{cld} \sim 0.6$), unfrozen land ($A_{lnd} \sim 0.2$), and snow and ice ($A_{ice} \sim 0.8$). The parameters $f_{xxx}$ in Eqn. 6 are the disk area fractions of each of the three surfaces and $A_{xxx}$ are their albedos.

We ignore the contribution of light that is backscattered from the ~50-meter deep photic zone, as well as Rayleigh-scattered light from the atmosphere that reflects from the ocean surface. Both of these sources are concentrated at short wavelengths (< 0.5 μm) and comprise < 6% of the light reflected from Earth's surface (Woolf *et al.* 2002). We now focus our attention on the fourth, specular term of Eqn. 6. The area of the planet that contributes to the specular signal is the fractional area of the planet covered by water $f_{wtr}A_{wtr}$ multiplied by the probability $p_{wav}$ of a wavy surface having the correct slope and azimuthal orientation for reflecting starlight in the direction of Earth. The albedo of seawater $A_{wtr} = \left( R_{\perp}^2 + R_{\parallel}^2 \right)/2$ , where $R_{\perp}$ and $R_{\parallel}$ are the classical Fresnel reflection coefficients for two polarization directions (Griffiths 1998), and depends strongly on illumination angle. Water is dark ($A \sim 0.04$) at zenith angles < 45º, but is mirror-like at angles approaching glancing incidence when its albedo climbs steeply toward 100%.

The relative strengths of the specular and diffuse reflections can be estimated from the sizes of the four bracketed terms in Eqn. 6. Supposing a planet disk to be a diurnally-averaged mix of unfrozen land, ice, and water in the proportion 25:20:55, and 2.5% of the ocean is covered by sea foam, then the diffuse terms of Eqn. 6 sum to

$$f_{foam}A_{foam} \quad + \quad f_{lnd}A_{lnd} \quad + \quad f_{ice}A_{ice}$$
$$(0.025)(0.6) \quad + \quad (0.25)(0.2) \quad + \quad (0.2)(0.8) \quad = 0.225$$

with 0% cloud cover, and

$$f_{foam}A_{foam} \quad + \quad f_{cld}A_{cld} \quad + \quad f_{lnd}A_{lnd} \quad + \quad f_{ice}A_{ice}$$
$$(0.012)(0.6) \quad + \quad (0.22)(0.6) \quad + \quad (0.12)(0.2) \quad + \quad (0.1)(0.8) \quad = 0.413$$

with 50% cloud cover.

Although 55% of our hypothetical planet disk is covered by water, only a tiny percentage of the ocean surface contributes to the specular term because the probability of waves being oriented properly for sending light in the direction of Earth is small; when the planet is in quadrature phase as in Fig. 3a, the disk-averaged value of $p_{wav}$ is found from the model to be $5.2 \times 10^{-6}$. Also, since starlight in this phase is incident the waves at small zenith angles, the average albedo of the ocean that is responsible for the glint is only ~4%. Thus the specular term in Eqn. 6 under clear skies is $(0.55)(5.2 \times 10^{-6})(0.04)/10^{-5} = 0.011$, or 5% of the diffuse signal.

However, both $p_{wav}$ and $A_{wtr}$ of Eqn. 6 increase rapidly with stellar zenith angle, which is large for planets in crescent phase with orbital inclinations near 90º (Fig. 3e and Fig. 4b). In this geometry, the ocean is obliquely illuminated and specular reflection is from small slopes on the ocean surface. Because small slopes occur with greater probability than large ones (Eqn. 3), the disk-averaged wave probability $p_{wav}$ increases to $1.3 \times 10^{-5}$, and the area of the glint spot is magnified (compare Fig. 3a and Fig 3e). In addition, the mean albedo of ocean water $A_{wtr}$ at glancing incidence grows to $> 0.6$, or more than 15 times the albedo at normal incidence. Thus, the specular term in Eqn. 6 becomes $(0.55)(1.3 \times 10^{-5})(0.6)/10^{-5} = 0.429$, or nearly twice the diffuse signal at this phase angle.

## Results and Discussion

Is the glint detectable in the disk-averaged signal from an Earth-like planet in crescent phase? To address this question, we first used the reflectance model to generate optical light curves for an Earth-like planet having an axial orientation corresponding to a disk-centred latitude of 23ºN and with an optimum (i = 90º) edge-on viewing geometry (Figs. 3e,f). Light curves are shown in Fig. 5a for an Earth with no clouds and with 50% clouds. The cloud-free light curve is naturally the fainter one and its shape cannot be modelled by an idealized Lambertian planet; matching the peak brightness or the slope at one phase is done at the expense of an acceptable fit at another phase. (Compare the lower Lambertian curve in Fig. 5a with the cloud-free model light curve.) A Lambertian light curve is naturally a better fit for the cloudy planet because of the isotropy of the reflected signal. The albedo of the 50% cloudy planet near full phase ($A$ = 0.35; read at $\theta$ = 180º) closely compares to the albedo of Earth ($A$ = 0.31), which lends support to our choice of albedos used for the individual surface types.

There are two key differences between the lower Lambertian curve and the cloud-free light curve. First, there is a significant asymmetry in the light curve about $\theta$ = 180º when the planet reaches peak illumination (were it not directly behind the star). The asymmetry is still evident, albeit less pronounced, in the cloudy-planet light curve, and stems from seasonal changes in surface albedo. The cloud-free Earth reaches peak brightness near $\theta$ = 150º (corresponding to the month of May) when the illuminated area of the northern hemisphere that is covered by snow and sea ice is at a maximum. The planet begins to dim well before peak illumination because of thawing and poleward retreat of snow and ice as the hemisphere warms. Identifying such an asymmetry in the light curve of an exoplanet could be evidence of a significant seasonal transformation on a planet surface, but it could also be a symptom of seasonal weather patterns and the associated ebb and flow of the clouds.

Both the ocean-covered planets with and without clouds are measurably brighter than the idealized planets near inferior conjunction ($\theta = 0°$ and $360°$). This is caused by the anisotropic reflection of starlight from the planet's oceans. In crescent phase, the light curve edges are elevated slightly by specular reflection of light off the ocean waves at glancing incidence. In both cases the planet begin to brighten relative to the Lambertian planets 40-60° before inferior conjunction, still within the edges of the observing window defined by the curves of Fig. 2. Even so, the specular edges of an illumination cycle will be the most difficult phases to observe because they occur when a planet is closest to its star and when it is faintest. For example, a crescent planet with 1/5th the illuminated area of a planet in quadrature will require an integration time that is $5^2 = 25$ × longer (or 14-21 days for TPF) to reach the same S/N ratio. (Compare the vertical error bars in Fig. 5a). Such long observations will be best performed on the largest terrestrial planets with relatively long orbital periods around G and F class stars.

Is the glint more pronounced on planets completely covered by water? To test the idea, we calculated the light curves of an Earth-like planet with 0% and 50% cloud cover as before but with the entire surface covered by water and without the seasonal albedo changes from snow and ice. Figure 5b shows that the uniform watery surface covered by a uniform cloud deck eliminates the seasonal asymmetry seen in Fig. 5a relative to the orbital conjunctions. The specular brightening in crescent phase that was evident in Fig 5a is slightly more pronounced, as expected, in this case.

More striking here is the approximately constant light curve in Fig. 5b for the cloud-free planet, which shows the planet to be just as bright in crescent phase as it is near full phase. This feature originates from our initial (but still valid) assumption that diffuse light reflected by the oceans is negligible in the spectral band $(0.6 - 1.0 \ \mu m)$ considered viable for studying extrasolar planets (Traub *et al.* 2006). Thus, most of the starlight incident the oceans is specularly reflected in directions away from the viewer, making a cloud-free water world appear extremely dark ($A_{wtr} \sim 0.02$). The light emanating from the glint spot dominates the observed signal at all phases. The combined measurement of an extremely low albedo coupled with the approximately level light curve would be evidence that a distant planet has surface water (and no clouds). We note, however, that real planets with variable, non-uniform cloud cover should oscillate between the cloud-free and partly-cloudy states shown in Figs. 5a and 5b on a diurnal-to-seasonal timescale. This will make identifying water from the light curve more difficult except on the chance occasion when the cloud cover diminishes and the planet darkens.

We now examine the influence of orbital inclination on the specular edges of the light curve. Fig. 2 shows that planets with inclinations $i < 30º$ never reach the crescent phase ($f_A < 0.25$) where the glint from water begins to dominate the disk-averaged signal. The glint becomes more pronounced at higher orbital inclinations as larger phase angles become observable. This trend is evident in Fig. 6a where we show the disk-averaged albedo as a function of orbital phase angle. [Astronomers should be able to determine the albedo of distant planets once planetary radius and orbital inclination are known.] For edge-on orbits, the albedo rises steeply at high phase angle near the edges of the observing window. The rise commences ~60º away from inferior conjunction and, thus, should be observable for at least 1/6 of the orbit, assuming the planet enters the glare of the star at θ ~ 30º. Adding clouds to the model naturally brightens the planet, but it does not eliminate or significantly reduce the glint at the edges of observation (dashed curve in Fig. 6a).

Similar brightening of the crescent Earth has been identified in earthshine reflected by the gibbous Moon (Pallé *et al.*; their Fig. 9). The earthshine data in Fig. 6a shows that the real Earth (observed between 1998 and 2002) is fainter and a better specular reflector than the model Earth under 50% cloud cover. This discrepancy cannot result from the real Earth having more water beneath the Moon at the time of observations compared to our diurnally-averaged model; We can cover the model Earth completely with water as before, thereby making the signal uniform over a rotation. Figure 6b shows the resulting albedo variation. Comparing the curves in Fig. 6b to Fig. 6a reveals there to be almost no difference in the phase angle of the specular upturn for Earth with continents compared to Earth without. This is because the contribution of the glint to the disk-averaged signal is more sensitive to phase angle than on the surface area of the glint spot.

Could the discrepancy between the model and earthshine data result from our neglecting anisotropic forward scattering by clouds? The shapes of plausible light-scattering phase functions for clouds make this scenario also unlikely. The double Henyey-Greenstein phase functions used by Hovenier and Hage (1989) to model the atmospheres of Earth, Jupiter, Venus, and Saturn do not begin rising in the forward direction until scattering angles fall below 60º (forward direction is 0º). This means that there is little forward scattering in an atmosphere until the angle between the star, scatterer, and observer is <

60º. The earthshine data in Fig. 6a and 6b indicates that the phase brightening of Earth begins earlier, near quadrature at ~270º (90º before conjunction).

Molecular scattering by the atmosphere might account for the early phase upturn in the earthshine signal compared to the model. Rayleigh-scattered light is concentrated at wavelengths < 0.5 µm and is not included in the model, but it could comprise a significant percentage of the light sent toward the gibbous moon with Earth in crescent phase. Detailed atmospheric modelling is needed to fully reconcile this problem; however, astronomers plan to observe extrasolar earths at wavelengths > 0.6-1.0 µm, where molecular-scattering is negligible. Future remote observations of Earth using interplanetary spacecraft will also help if the Earth can be observed at extreme phase angles, such as by spacecraft in orbit around or en route to Mars.

Earthshine has also been observed to be strongly polarized (Coffeen 1979; their Fig. 14), peaking at 40% linear polarization near quadrature. The source of the polarization is scattering by clouds, molecular scattering by the atmosphere, and reflection from the surface. Satellite observations of Earth clouds, as well as spacecraft observations of Venus and Jupiter (Ibid), show that clouds can polarize a signal by 10-20%. However, clouds observed on the limbs of these planets have yielded polarization percentages as high as 75% from the sunlight being scattered at glancing incidence. Recent modeling by Stam *et al.* (2006) has demonstrated that it will be worthwhile to extend such polarimetric measurements to extrasolar giant planets, and McCullough (*in review*) has given some attention to the polarimetry of Earth and Earth-like planets, as we also do here.

Our final goal for this paper is to use our model to see how large a polarization percentage is possible from reflection of light off a planet surface with water. Reflected light is partitioned into two states of polarization $F_\perp$ and $F_\parallel$ using Eqn. 6 and the degree of polarization $P$ is found from

$$P = \frac{(F_\perp - F_\parallel)}{(F_\perp + F_\parallel)} \qquad (7)$$

Diffuse light is divided evenly between $F_\perp$ and $F_\parallel$, and both states of polarization are averaged over the planet disk and over a complete rotation, as earlier. We also set orbital inclination $i = 90°$ to reach the maximum disk-averaged polarization.

Figure 7a shows how the percentage $P$ varies around an orbit for a planet with Earth geography and no clouds, and Fig. 7b shows the same for a planet covered completely by water. Both figures show the most significant polarization from surface reflection in crescent phase. The peak polarization for the entirely water-covered Earth-like planet is more than twice the percentage for Earth with continents and occurs at a slightly earlier phase angle. However, neither model reaches maximum polarization where Earth does (near quadrature, $\theta = 90°$), possibly because we here ignore the contribution of atmospheric Rayleigh scattering. A second set of computations with 50% cloud cover (not shown) give peak polarizations $P \sim 10\text{-}15\%$, which are comparable to the values observed for the cloudy atmospheres of Venus and Jupiter.

The above result suggests that it will extremely difficult to distinguish between atmospheric and surface reflection from a polarimetric light curve. However, a joint analysis of the photometric and polarimetric data will help to remove some of the ambiguity and possibly constrain the nature of a planet surface. For example, the polarization peaks in Fig. 7a and Fig. 7b could be cited as evidence of surface water, but they might also be produced by clouds. However, photometry shows the disk of the model planet to be darker than any cloudy world in the Solar System. Thus, the strength of the polarized signal combined with a dark, non-Lambertian photometric light curve would together serve as evidence that the planet has surface water.

## Summary

Half of all extrasolar planets will have orbital inclinations in the range $60° < i < 120°$ where glint from surface water is in principle detectable. Specular reflection of starlight from an ocean surface occurs at all phase angles, but only begins to dominate the whole-disk signal when a planet is nearest its star as a thin crescent. Observations at such phase angles can be obtained of planets around G and F stars where they have adequate angular separation and orbit within the habitable zone. Planets with large oceans will scatter light non-uniformly and exhibit non-Lambertian photometric light curves. Such planets will also appear markedly dark, with surface albedos in the range 5-15%, under average illumination. Significant polarization of the reflected beam in crescent phase will further indicate that water is present on a planet's surface. This suggests that of all the extremely difficult measurements astronomers hope to make with a TPF-class telescope, time-series photometry and polarimetry that can lead to the identification of specular reflection from surface water might be the easiest.

# Acknowledgements

We thank E. Pallé for helpful discussions regarding the earthshine measurements and V.I. Haltrin for help with the wave-reflection algorithm. We also thank E. Ford for a thoughtful and valuable review. This work was supported by the NASA Terrestrial Planet Finder Foundation Science program by grant NNG04GL48G and the NASA Astrobiology Institute through cooperative agreement NNA04CC08A.

Correspondence and requests for materials should be addressed to D.W. (dmw145@psu.edu).

# Figure Captions

**Fig. 1** Time needed to observe an Earth-sized planet above an acceptable S/N using a future TPF-C class telescope. Orbital inclination $i$ (here set to 90°) determines the illuminated fraction of the viewing disk $f_A = (1- \cos\theta \sin i)/2$. Orbital longitude $\theta$ is 0° and 360° when the planet transits the star and is 180° when it is behind the star. The time needed to reach S/N = 10 for an Earth-sized planet in quadrature ($\theta = 90°$) of its phase cycle around a top-priority TPF-C target star is ~2 days according to Brown (2005). Also, S/N $\propto \tau^{1/2} f_A$ since the signal grows as $\tau f_A$ and the noise is $\sim\tau^{1/2}$. According to the above figure, TPF-C should be able to image a crescent exo-earth up to S/N = 3 in 5-15 days. Vertical gray zones denote regions where a planet would be too close to its star to resolve (assuming an inner working angle ~4$\lambda$/D) with a 3×8 meter telescope from a distance of ~10 parsecs.

**Fig. 2** Phase cycles of planets as a function of orbital inclination *i* and orbital longitude

θ. Thick black lines denote regions where a planet is a thin crescent ($f_A < 0.25$), and the

lines are truncated where a planet orbiting at 1AU from its star reaches the inner

working angle ~4$\lambda$/D = 0.057'' (marked by the edge of the vertical gray zones for *i* =

90°). Limiting phases for other star-planet separations are marked in the lower right

corner of the diagram. For example, a planet at 2.0 AU from its star can be resolved

(assuming *i* = 90°) down to an extremely narrow crescent corresponding to $f_A$ ~0.03,

compared to $f_A$ ~0.09 at 1.0 AU. Dotted lines mark regions of the phase cycle when the planet cannot be resolved. Earth diagrams above the figure show the phases.

**Fig. 3** Model earths seen at different viewing orientations. Oceanic glint is shown as an anomalously bright white patch on the dark gray ocean surface. Orbital inclination $i$ is the angle between the orbit normal and the observer direction. Orbital longitude $\theta = 0°$ and $360°$ at inferior conjunction and $180°$ at superior conjunction with the star between the observer and the planet. Model planet spin axes are inclined $23.5°$ relative to their orbital planes, as for the Earth, and are arbitrarily oriented so that northern hemisphere winter solstice occurs at inferior conjunction. Month labels show which of the twelve surface albedo maps are in view. Model planets are on circular orbits and have Earth-like orbital periods and day lengths. Reflected light is averaged over the entire viewing disk and over a complete rotation.

**Fig. 4** Views of the gibbous Earth (a) and crescent Earth (b) from space. The gibbous Earth photo (Courtesy NASA *Blue Marble* Project: visibleearth.nasa.gov) is a composite satellite image of Earth showing the northern ice cap, abundant cloud cover, and a glint of sunlight off the eastern Pacific Ocean (marked with an orange circle). The crescent Earth was imaged by the MESSENGER spacecraft (messenger.jhuapl.edu) on its way to Mercury on 2 Aug. 2005 and shows a similar glint just west of the Galapagos and South America.

**Fig. 5** Orbital light curves of the model Earth compared to an idealized Lambertian planet. Apparent brightness is the product $f_A$ [ ], where [ ] includes the bracketed terms of Eqn. 6. The model light curves (thick black line - 0% clouds: dashed line - 50% clouds) are smoothed fits to the results of 72 diurnally-averaged, whole-disk integrations of reflected light. Here, the orbit is viewed edge-on and the disk-centered latitude is the Tropic of Cancer. Earth geography is used in panel **a**, and the surface is uniformly covered with water in panel **b**. The pair of Lambertian curves (thin gray lines) were obtained using Eqn. 1, and their amplitudes were adjusted to give acceptable fits to the model. Vertical "error bars" show the approximate integration times needed to observe planets of different apparent brightness.

**Fig. 6** Albedo variation of the model Earth seen at different orbital inclinations. Here we plot the sum of area-weighted albedo terms in brackets [ ] in Eqn. 6, also equal to the apparent brightness of Fig. 5 normalized by the illuminated area fraction $f_A$. Earth geography is used in panel **a**, and an all-water surface is used in panel **b**. The case of Earth with 50% cloud cover and 90° inclination is indicated with a dashed line. Earthshine data are from Pallé *et al.*; Fig. 9.

**Fig. 7** Orbital variation in polarization percentage (Eqn. 7) of the disk-averaged signal reflected from the surface of an Earth-like planet (thick black line). Earth geography is used in panel **a**, and an all-water surface is used in panel **b**. For both panels, orbital inclination is set to 90º and cloud cover is set to 0%. Light curves for the two orthogonal states of polarization $F_\perp$ and $F_\parallel$, are also shown using dotted and dashed lines, and a thin solid line is used for the combined flux. Fluxes are plotted using the scale on the alternate y-axis. For the all-water planet in panel **b**, fluxes are magnified 10× for clarity.

**Figure 1**

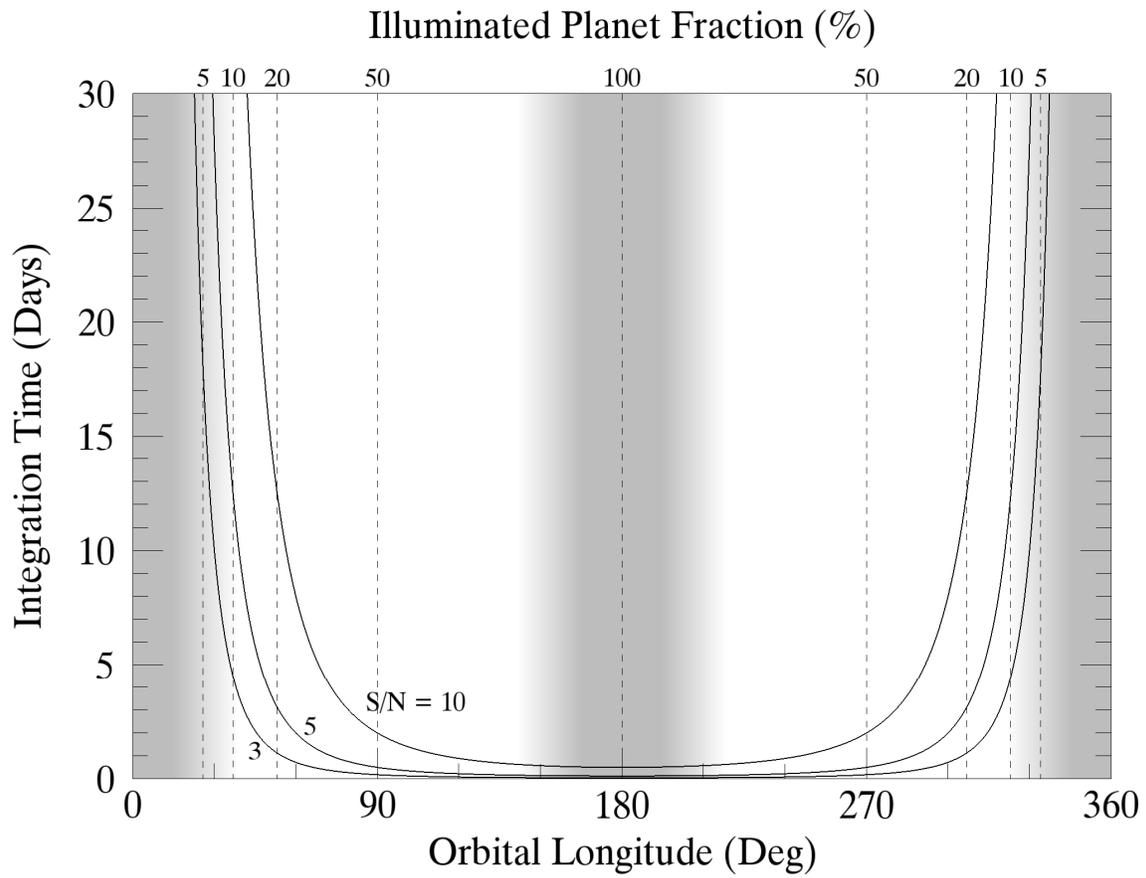

**Figure 2**

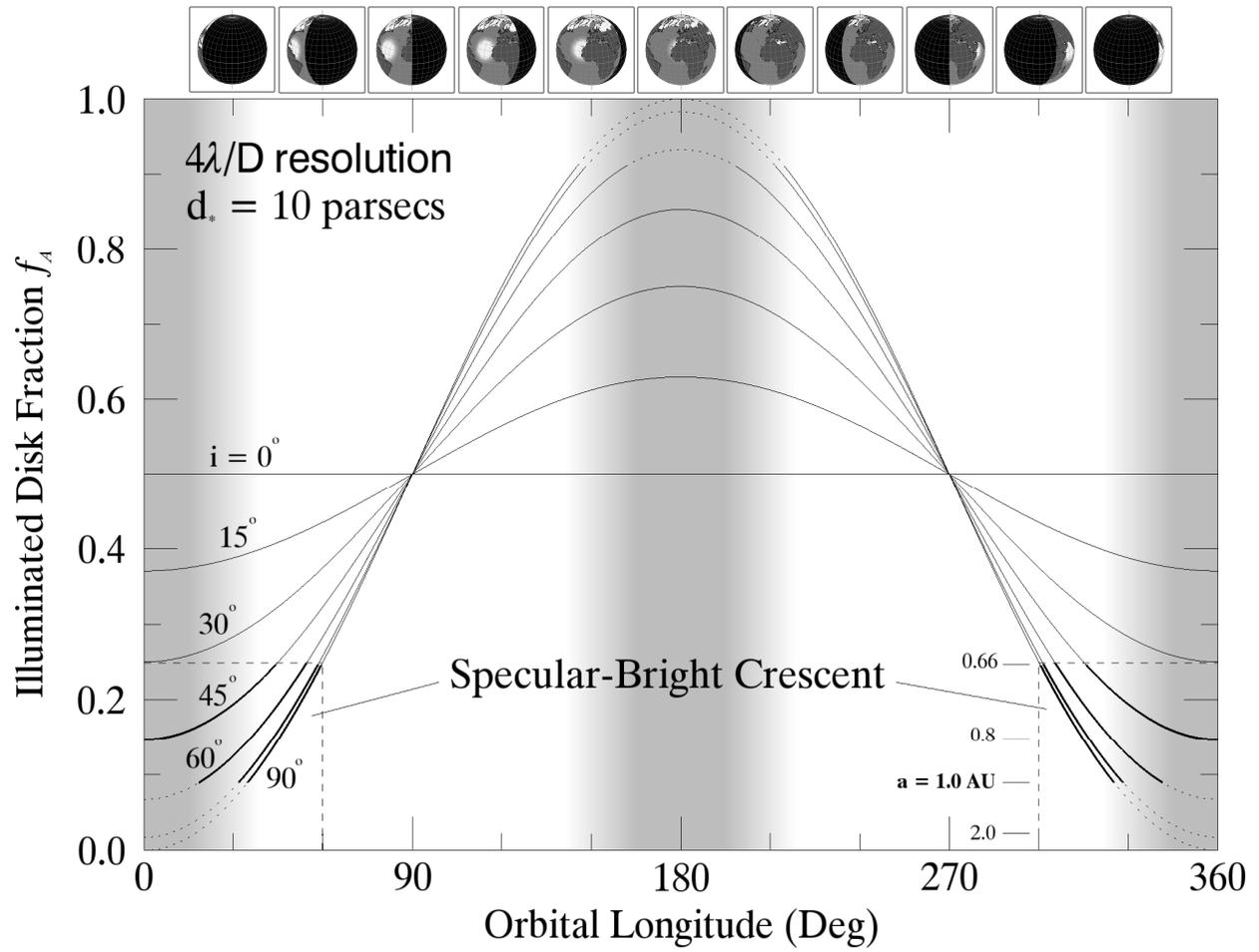

**Figure 3**

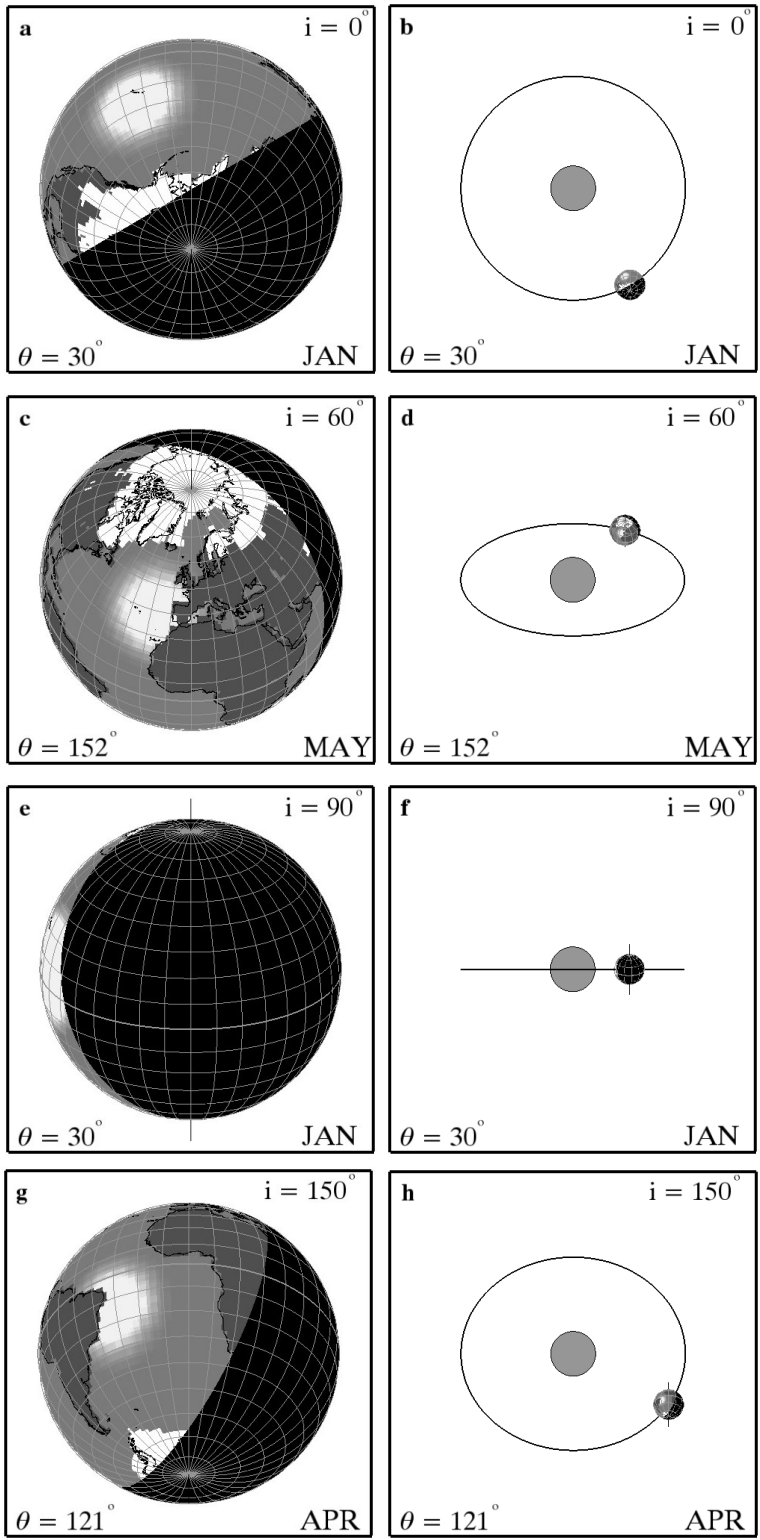

**Figure 4**

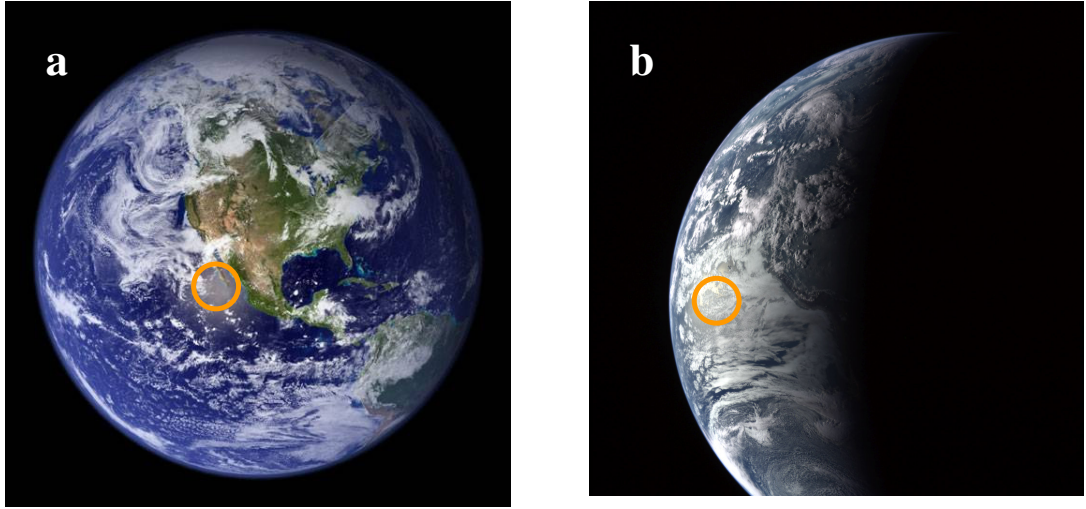

**Figure 5a**

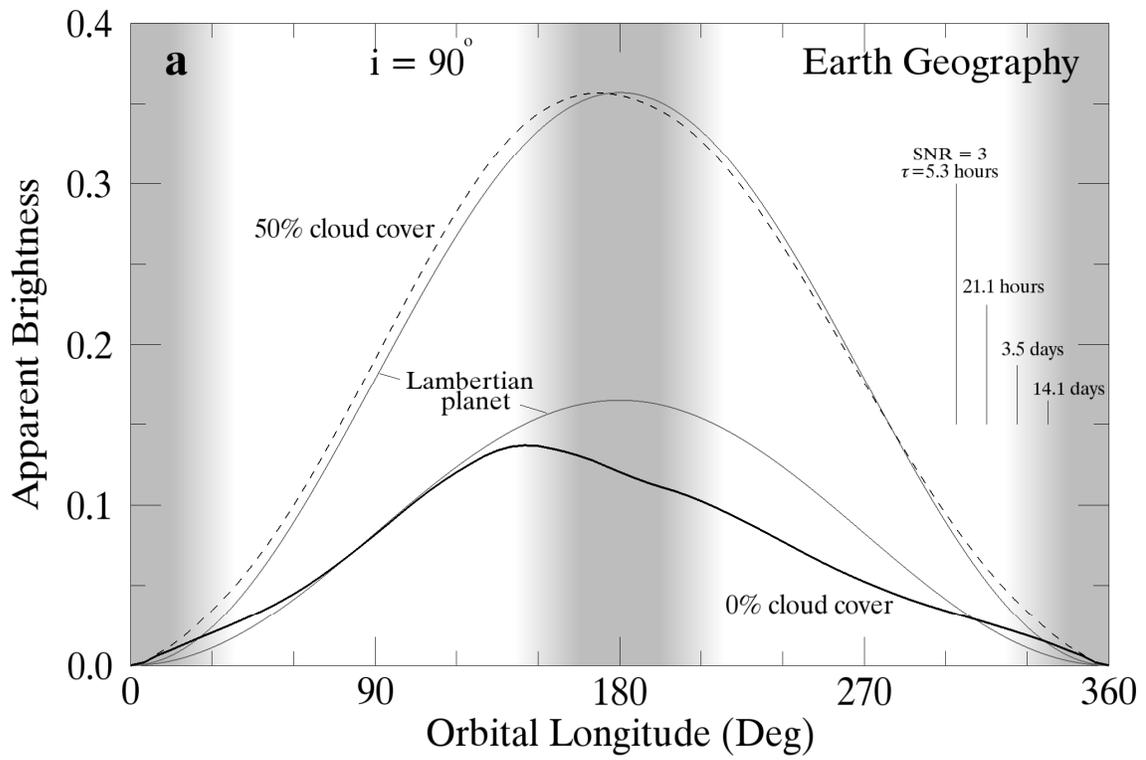

**Figure 5b**

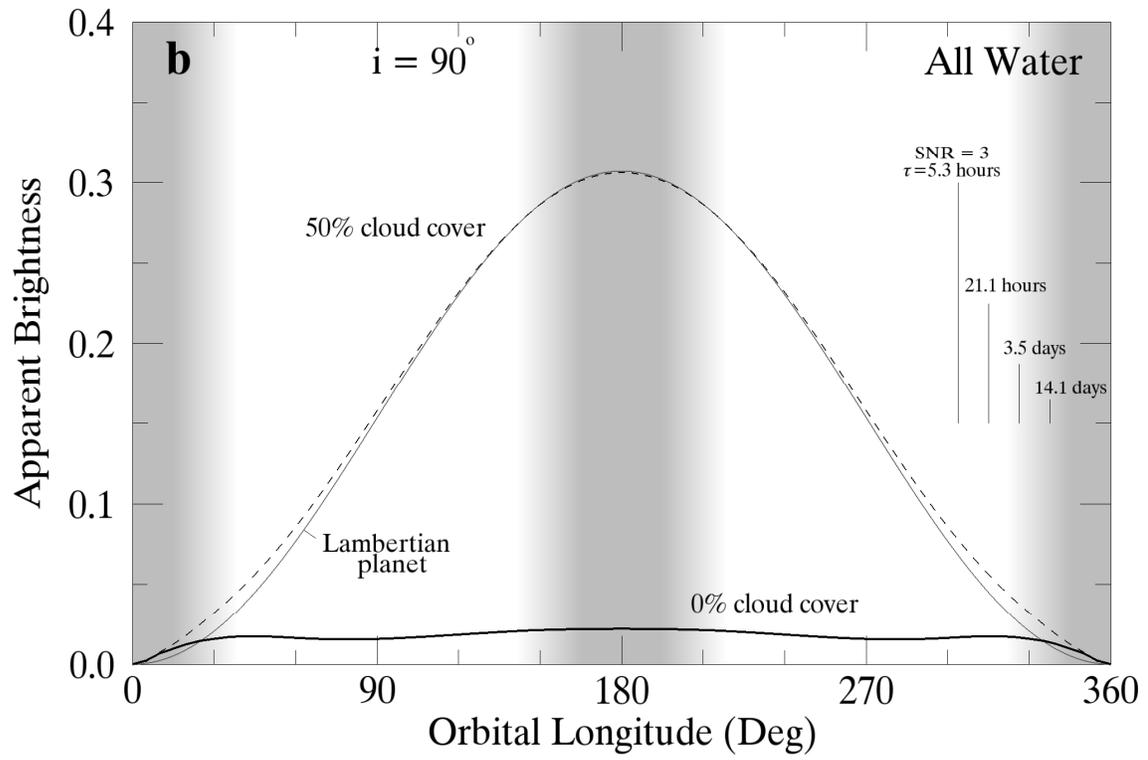

**Figure 6a**

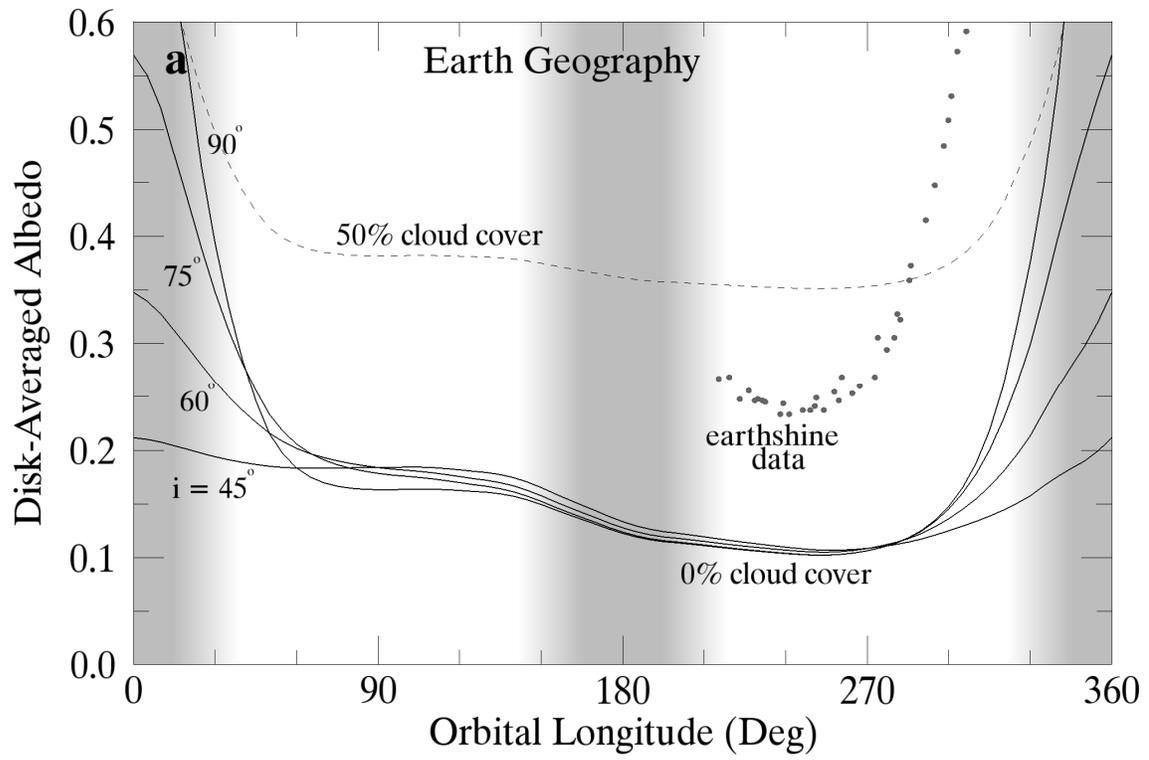

**Figure 6b**

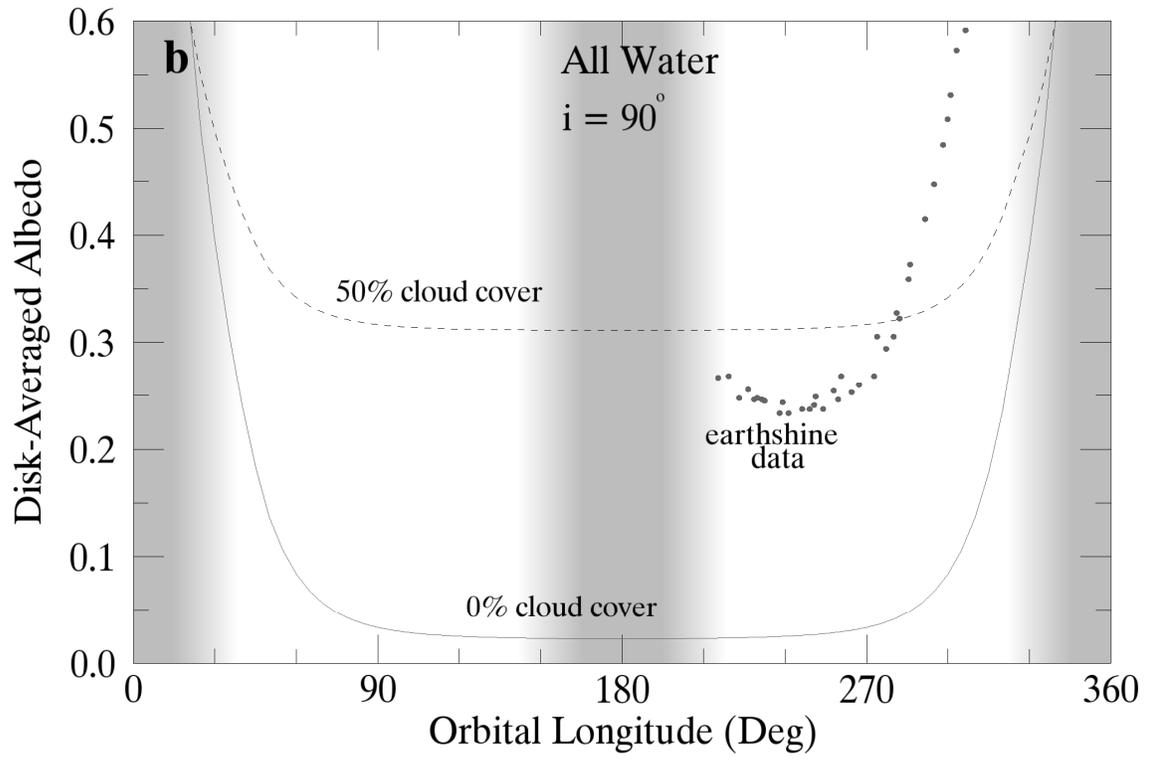

**Figure 7a**

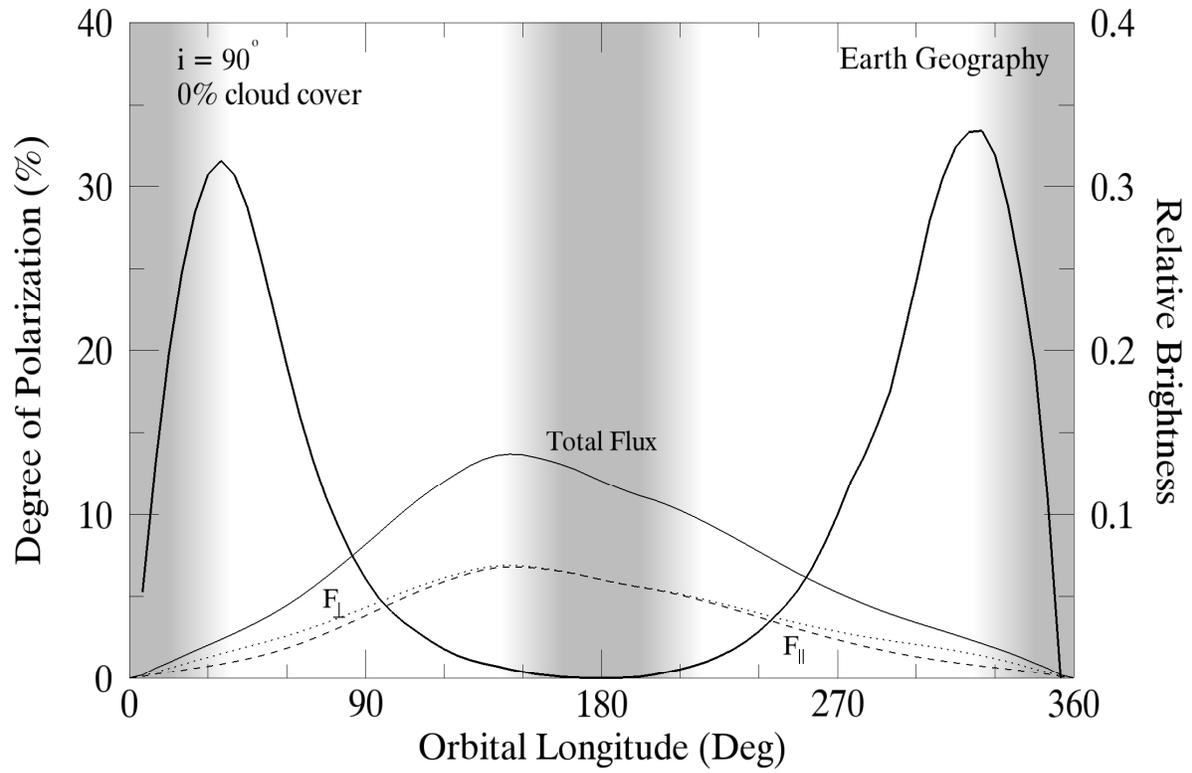

**Figure 7b**

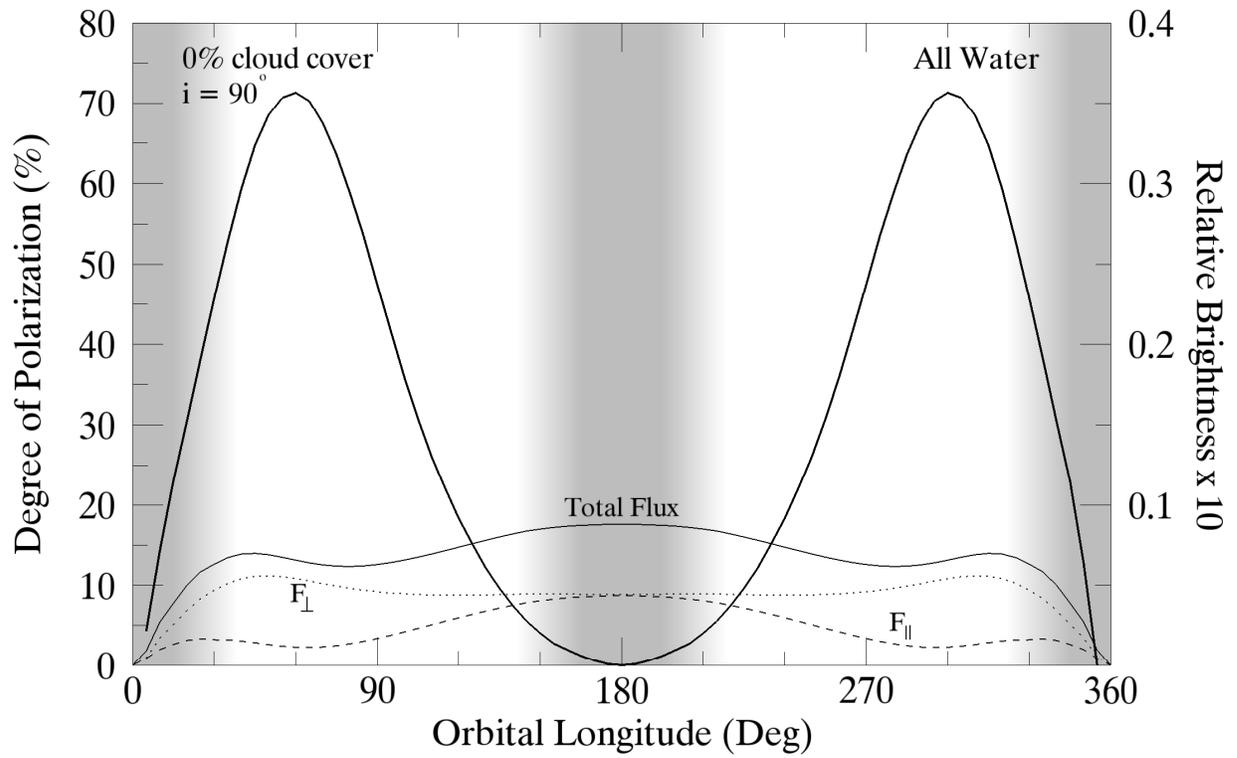